\documentclass[twocolumn]{aastex631}

\usepackage{color}

\newcommand {\swift} {\textsl{Swift}}
\newcommand {\fermi} {\textsl{Fermi}}
\newcommand {\nustar} {\textsl{NuSTAR}}
\newcommand {\nicer} {\textsl{NICER}}
\newcommand {\parkes} {{Parkes}}
\newcommand {\src} {1E~1547.0$-$5408}

\def \arcmin {\hbox{$^\prime$}}
\def \arcsec {\hbox{$^{\prime\prime}$}}

\shorttitle{2022 outburst of 1E~1547.0$-$5408}
\shortauthors{Lower et al.}

\graphicspath{{./}{figures/}}

\begin{document}

\title{The 2022 high-energy outburst and radio disappearing act of the magnetar 1E~1547.0$-$5408}

\correspondingauthor{Marcus E. Lower, George Younes}
\email{marcus.lower@csiro.au, george.a.younes@nasa.gov}

\author[0000-0001-9208-0009]{Marcus E. Lower}
\affiliation{Australia Telescope National Facility, CSIRO, Space and Astronomy, PO Box 76, Epping, NSW 1710, Australia}

\author[0000-0002-7991-028X]{George Younes}
\affiliation{Astrophysics Science Division, NASA Goddard Space Flight Center, Greenbelt, MD 20771, USA}
\affiliation{Department of Physics, The George Washington University, 725 21st Street NW, Washington, DC 20052, USA}

\author[0000-0002-7374-7119]{Paul Scholz}
\affiliation{Dunlap Institute for Astronomy \& Astrophysics, University of Toronto, 50 St. George Street, Toronto, ON M5S 3H4, Canada}

\author[0000-0002-1873-3718]{Fernando Camilo}
\affiliation{South African Radio Astronomy Observatory, 2 Fir Street, Black River Park, Observatory 7925, South Africa}

\author[0000-0002-1769-6097]{Liam Dunn}
\affiliation{School of Physics, University of Melbourne, Parkville, VIC 3010, Australia}
\affiliation{OzGrav: The ARC Centre of Excellence for Gravitational-wave Discovery, Parkville, VIC 3010, Australia}

\author[0000-0002-7122-4963]{Simon Johnston}
\affiliation{Australia Telescope National Facility, CSIRO, Space and Astronomy, PO Box 76, Epping, NSW 1710, Australia}

\author[0000-0003-1244-3100]{Teruaki Enoto}
\affiliation{RIKEN Cluster for Pioneering Research, 2-1 Hirosawa, Wako, Saitama 351-0198, Japan}

\author{John M. Sarkissian}
\affiliation{Australia Telescope National Facility, CSIRO, Space and Astronomy, Parkes Observatory, PO Box 276, Parkes NSW 2870, Australia}

\author{John E. Reynolds}
\affiliation{Australia Telescope National Facility, CSIRO, Space and Astronomy, PO Box 76, Epping, NSW 1710, Australia}

\author[0000-0001-7128-0802]{David M. Palmer}
\affiliation{Los Alamos National Laboratory, Los Alamos, NM, 87545, USA}

\author{Zaven Arzoumanian}
\affiliation{Astrophysics Science Division, NASA Goddard Space Flight Center, Greenbelt, MD 20771, USA}

\author[0000-0003-4433-1365]{Matthew G. Baring}
\affiliation{Department of Physics and Astronomy - MS 108, Rice University, 6100 Main Street, Houston, Texas 77251-1892, USA}

\author[0000-0001-7115-2819]{Keith Gendreau}
\affiliation{Astrophysics Science Division, NASA Goddard Space Flight Center, Greenbelt, MD 20771, USA}

\author[0000-0002-5274-6790]{Ersin G\"o\u{g}\"u\c{s}}
\affiliation{Sabanc\i~University, Faculty of Engineering and Natural Sciences, \.Istanbul 34956 Turkey}

\author[0000-0002-6449-106X]{Sebastien~Guillot}
\affil{IRAP, CNRS, 9 avenue du Colonel Roche, BP 44346, F-31028 Toulouse Cedex 4, France}
\affil{Universit\'{e} de Toulouse, CNES, UPS-OMP, F-31028 Toulouse, France}

\author[0000-0001-9149-6707]{Alexander J. van~der~Horst}
\affiliation{Department of Physics, The George Washington University, 725 21st Street NW, Washington, DC 20052, USA}

\author[0000-0001-8551-2002]{Chin-Ping Hu}
\affiliation{Department of Physics, National Changhua University of Education, Changhua 50007, Taiwan}

\author[0000-0003-1443-593X]{Chryssa Kouveliotou}
\affiliation{Department of Physics, The George Washington University, 725 21st Street NW, Washington, DC 20052, USA}

\author[0000-0002-0633-5325]{Lin Lin}
\affiliation{Department of Astronomy, Beijing Normal University, Beijing 100875, China}

\author[0000-0002-0380-0041]{Christian~Malacaria}
\affiliation{International Space Science Institute (ISSI), Hallerstrasse 6, 3012 Bern, Switzerland}

\author[0000-0002-0254-5915]{Rachael Stewart}
\affiliation{Department of Physics, The George Washington University, 725 21st Street NW, Washington, DC 20052, USA}

\author[0000-0002-9249-0515]{Zorawar Wadiasingh}
\affiliation{Astrophysics Science Division, NASA Goddard Space Flight Center, Greenbelt, MD 20771, USA}
\affiliation{Department of Astronomy, University of Maryland, College Park, MD 20742, USA}
\affiliation{Center for Research and Exploration in Space Science and Technology, NASA/GSFC, Greenbelt, Maryland 20771, USA}

\begin{abstract}

We report the radio and high-energy properties of a new outburst from the radio-loud magnetar \src. Following the detection of a short burst from the source with \swift-BAT on 2022 April 7, observations by \nicer\ detected an increased flux peaking at $(6.0 \pm 0.4) \times 10^{-11}$\,erg\,s$^{-1}$\,cm$^{-2}$ in the soft X-ray band, falling to the baseline level of $1.7\times10^{-11}$\,erg\,s$^{-1}$\,cm$^{-2}$ over a 17-day period.
Joint spectroscopic measurements by \nicer\ and \nustar\ indicated no change in the hard non-thermal tail despite the prominent increase in soft X-rays. 
Observations at radio wavelengths with \textsl{Murriyang}, the 64-m Parkes radio telescope, revealed that the persistent radio emission from the magnetar disappeared at least 22 days prior to the initial \swift-BAT detection and was re-detected two weeks later.
Such behavior is unprecedented in a radio-loud magnetar, and may point to an unnoticed slow rise in the high-energy activity prior to the detected short-bursts.
Finally, our combined radio and X-ray timing revealed the outburst coincided with a spin-up glitch, where the spin-frequency and spin-down rate increased by $0.2 \pm 0.1$\,$\mu$Hz and $(-2.4 \pm 0.1) \times 10^{-12}$\,s$^{-2}$ respectively. 
A linear increase in spin-down rate of $(-2.0 \pm 0.1) \times 10^{-19}$\,s$^{-3}$ was also observed over 147\,d of post-outburst timing. 
Our results suggest that the outburst may have been associated with a reconfiguration of the quasi-polar field lines, likely signalling a changing twist, accompanied by spatially broader heating of the surface and a brief quenching of the radio signal, yet without any measurable impact on the hard X-ray properties.

\end{abstract}

\keywords{High-energy astrophysics (739) --- Magnetars (992) --- Neutron stars (1108) --- Pulsars (1306) --- Radio pulsars (1353)}

\section{Introduction} \label{sec:intro}

Magnetars are slowly rotating neutron stars with the strongest known magnetic fields in the Universe.
Outbursts from these highly-magnetized objects are typically manifested by the sudden enhancement of their persistent X-ray emission by up to three orders of magnitude, which is often accentuated by spectral hardening and increased levels of complexity in their X-ray pulse profiles (see \citealt{CotiZelati2018} for a review).
This increase in high-energy activity is sometimes accompanied by the emission of up to hundreds of short-duration, bright hard X-ray bursts.
Five out of the nearly 30 known magnetars have also been found to display pulsed radio emission during outburst states \citep{Olausen2014}\footnote{\url{https://www.physics.mcgill.ca/~pulsar/magnetar/main.html}}. 
These radio-loud episodes are typically preceded by several years of radio quiescence followed by the sudden activation of intense, pulsed radio emission. 
Strong temporal variability in the detected radio flux, pulse-profile shape and polarization fraction are often observed \citep{Levin2012, Camilo2018, Dai2019, Lower2021a}. Moreover, rotational variations in the form of rapid increases in spin-down and sudden spin-up (glitches) or, less commonly, spin-down events (anti-glitches) are also detected following outbursts \citep{Archibald2013, Scholz2017, Younes2020, Caleb2022, hu20ApJ}.  These observational properties are likely a result of magnetic stresses from the decaying internal magnetic fields precipitating sudden crustal motion or fracturing, leading to an impulsive release of energy in the form of electromagnetic radiation and occasionally an alteration of the external magnetic field topology \citep{Duncan1992, Harding1999, Thompson2002, younes22ApJL18301}

With a spin period of $\sim 2.1$\,s, 1E~1547.0$-$5408 (also known as PSR~J1550$-$5418) is the second fastest rotating magnetar found to date.
First discovered as a persistent X-ray source by the \textit{Einstein Observatory} \citep{Lamb1981}, it was not until several decades later that it was recognized as a potential magnetar candidate embedded within the putative supernova remnant G327.24$-$0.13, estimated to be $\sim 4$\,kpc away \citep{Gelfand2007}.
Targeted pulsar-search observations revealed the presence of highly-polarized, radio pulses originating from the source \citep{Camilo2007}.
Subsequent timing observations enabled measurements of both the aforementioned spin-period and a spin-down rate of $2.138 \times 10^{-11}$\,s\,s$^{-1}$, thereby confirming its magnetar status (with an equatorial field of $B \approx 2\times 10^{14}$\,G; \citealt{Camilo2007}) and making it the second radio-loud magnetar to be discovered \citep{Camilo2007, Camilo2008}.
This serendipitous discovery appeared to follow an unnoticed outburst sometime in the mid-2000s \citep{Camilo2008} (possibly during 2007; see \citealt{Halpern2008}), and was soon followed by two intense high-energy outbursts in 2008 and 2009 \citep{Israel2010, Mereghetti2009, Scholz2011, Bernardini2011, vanderHorst12ApJ}.
Detections of the radio pulses became highly intermittent following the outbursts, but the source stabilized to a persistent radio-loud state after several years (Camilo et al. in prep.).

Since its major outburst of 2009 January, \src\ has remained in an unusual high X-ray flux state, bearing a much stronger resemblance to a persistent magnetar as opposed to its initial `transient' classification (\citealt{CotiZelati2020}; Camilo et al. in prep.).
No additional X-ray outbursts above the baseline were observed during this time, barring the detection of two short bursts seen with the the Neil Gehrels \swift\ Observatory (thereon Swift) Burst Alert Telescope (BAT) in 2011 and 2013, respectively \citep{2013GCN.14911....1H}.
Neither of these two short bursts were followed up by targeted high-cadence monitoring.

In this work, we present the detection of a new outburst from \src\ that began between late-March and early-April 2022.
An overview of our radio and high-energy observations are detailed in the following section.
In Section~\ref{sec:analysis} we present our analysis of the timing and radiative properties of \src\ during the outburst.
We conclude in Section~\ref{sec:disc} with a discussion on the interplay between changes in the high-energy, radio, and timing properties of \src, and potential implications for the outburst mechanism.

\section{Observations} \label{sec:obs}

\subsection{\parkes\ Observatory}

Regular monitoring of \src\ with an approximately 10-day cadence has been undertaken with \textsl{Murriyang}, the 64-m \parkes\ radio telescope, under the P885 project since its discovery as a radio-loud magnetar in 2007 \citep{Camilo2007}.
Our observations from 2019 January have been performed using the Ultra-Wideband Low (UWL) receiver system \citep{Hobbs2020}, where two independent pulsar search-mode data streams are recorded with the pulsar digital filterbank v4 (PDFB4) and GPU-based {\sc medusa} signal processors.
Individual observations lasted between 2--15\,minutes. 
For this work, we analyzed the subset of the PDFB4 data spanning from 2021 June to 2022 August and {\sc medusa} data covering 2021 December to 2022 August.
Data collected by PDFB4 were taken with 1\,ms time-sampling at a center frequency of 3100\,MHz with 512 channels covering 1024\,MHz of bandwidth.
Full Stokes data collected with {\sc medusa} were recorded with 128\,$\mu$s time resolution with 3328 channels covering the full UWL band from 704--4032\,MHz, and coherently dedispersed at a dispersion measure (DM) of 830\,pc\,cm$^{-3}$.
Both data sets were subsequently folded at the rotation period of the magnetar using {\tt dspsr} \citep{vanStraten2011}, such that a single pulse period is covered by 1024 phase bins. 
Frequency channels affected by radio-frequency interference (RFI) were excised using the {\tt paz} and {\tt pazi} tools in {\tt psrchive} \citep{Hotan2004, vanStraten2012}.
The folded {\sc medusa} data were then flux and polarization calibrated following the procedures outlined in \citet{Lower2020}. 

\subsection{\nicer}

The Neutron star Interior Composition ExploreR (\nicer) is a non-imaging, soft X-ray telescope mounted on the International Space Station, with a $\approx5\arcmin$ diameter field-of-view (FoV; \citealt{gendreau16SPIE}). 
\nicer\ provides a large collective area peaking at $1900$~cm$^2$ at 1.5~keV and a timing resolution $<300$\,ns. We initiated a series of \nicer\ target of opportunity observations of \src\ two days after the 2022 April 7 \swift-BAT detection of a magnetar-like burst coincident with \src. The \nicer\ monitoring consisted of about 1~ks observations every few days, starting on 2022 April 9. We provide the observations log in Table~\ref{xObsLog}.

\begin{table}[t]
\centering 
\caption{List of X-ray observations analyzed in this work}
\label{xObsLog}
\newcommand\T{\rule{0pt}{2.6ex}}
\newcommand\B{\rule[-1.2ex]{0pt}{0pt}}
\begin{center}{
\resizebox{0.47\textwidth}{!}{
\begin{tabular}{l c c c}
\hline
\hline
Telescope \T\B & Observation ID & Date & Total GTI \\

          \T\B &  & (UTC) & (ks) \\
\nustar\ & 30101035002 & 2016-04-23 & 84 \\
\nustar\ & 30401008002 & 2019-02-15 & 90 \\
\swift/XRT & 00013176022 & 2022-03-05 & 2.6 \\
\swift/XRT & 00013176023 & 2022-03-06 & 1.8 \\
\nicer\ & 5020300101 & 2022-04-09 & 1.9 \\
\nicer\ & 5020300102 & 2022-04-12 & 0.1 \\
\nicer\ & 5020300103 & 2022-04-13 & 0.2 \\
\nicer\ & 5020300104 & 2022-04-14 & 0.3 \\
\nicer\ & 5020300106 & 2022-04-16 & 1.2 \\
\nicer\ & 5020300107 & 2022-04-17 & 0.6 \\
\nicer\ & 5020300108 & 2022-04-18 & 0.4 \\
\nicer\ & 5020300109 & 2022-04-21 & 1.5 \\
\nustar\ & 90801309002 & 2022-04-21 & 24 \\
\nicer\ & 5020300110 & 2022-04-22 & 1.8 \\
\nicer\ & 5020300111 & 2022-04-25 & 0.4 \\
\swift/XRT & 00013176025 & 2022-05-07 & 3.0 \\
\nicer\ & 5020300113 & 2022-05-08 & 1.6 \\
\swift/XRT & 00013176026 & 2022-05-11 & 1.5 \\
\nicer\ & 5020300114 & 2022-05-13 & 0.9 \\
\nicer\ & 5020300115 & 2022-05-15 & 0.2 \\
\nicer\ & 5020300116 & 2022-05-21 & 1.0 \\
\nicer\ & 5020300117 & 2022-05-22 & 0.5 \\
\nicer\ & 5020300118 & 2022-05-28 & 0.7 \\
\nicer\ & 5020300119 & 2022-06-07 & 1.5 \\
\nicer\ & 5020300120 & 2022-06-17 & 1.2 \\
\nicer\ & 5020300121 & 2022-06-27 & 0.4 \\
\nicer\ & 5020300123 & 2022-07-07 & 0.9 \\
\swift/XRT & 00013176027 & 2022-07-09 & 4.3 \\
\nicer\ & 5020300124 & 2022-07-13 & 0.4 \\
\nicer\ & 5020300125 & 2022-07-17 & 2.1 \\
\nicer\ & 5020300126 & 2022-07-20 & 1.1 \\
\nicer\ & 5020300127 & 2022-07-21 & 0.7 \\
\nicer\ & 5020300128 & 2022-07-28 & 1.3 \\
\nicer\ & 5020300129 & 2022-07-30 & 0.4 \\
\nicer\ & 5020300130 & 2022-08-02 & 0.1 \\
\nicer\ & 5020300131 & 2022-08-03 & 0.2 \\
\nicer\ & 5020300132 & 2022-08-05 & 0.2 \\
\nicer\ & 5020300133 & 2022-08-09 & 0.4 \\
\nicer\ & 5020300134 & 2022-08-15 & 1.8 \\
\nicer\ & 5020300135 & 2022-08-17 & 2.3 \\
\hline
\hline
\end{tabular}}}
\end{center}
\end{table}


We cleaned and calibrated all \nicer\ data utilizing {\tt NICERDAS} version v008c as part of {\tt HEASoft} version 6.30.1. We applied the standard filtering criteria as described in the \nicer\ Data Analysis Guide\footnote{\href{https://heasarc.gsfc.nasa.gov/docs/nicer/data\_analysis/nicer\_analysis\_guide.html}{NICER data analysis guide.}}. Moreover, detectors 14 and 34 can show larger than normal background noise, hence, were excluded from all analyses. Finally, we compared the light curve (binned with a 5~second resolution) created in the energy range 1--6~keV (where the source is supposed to dominate the emission) to one created in the 13--15~keV energy range. Almost all the counts in the latter band are due to high energy particle background. We eliminated any simultaneous flaring background intervals that appeared in both light curves. These were mostly present at the start and/or end of a good time interval (GTI) due to imprecise modelling of the South Atlantic Anomaly. We provide the total GTIs of each \nicer\ observation in Table~\ref{xObsLog}. We utilise {\tt Xselect} as part of {\tt HEASoft} version 6.30.1 to extract the source spectra, and estimated the background contribution for each observation utilizing the 3C50 model \citep{remillard2021:3c50}.

\subsection{\nustar}

\nustar\ (Nuclear Spectroscopic Telescope ARray; \citealt{harrison13ApJ:NuSTAR}) is a focusing hard (3--79~keV) X-ray telescope, consisting of two identical focal plane modules, FPMA and FPMB. Each module is formed by an array of $2\times2$ detectors providing a total of about $12\arcmin\times12\arcmin$ FoV. Prompted by the increase in the soft X-ray flux, we requested a \nustar\ director discretionary time and \src\ was observed for 24~ks on 2022 April 21. We utilized {\tt NuSTARDAS} software version 2.1.2 to clean and calibrate the event files. We employed the task {\tt nuproducts} to extract source events, light curves, and spectra from a circular region with a 50\arcsec-radius around the source central brightest pixel, which corresponds to 70\% of the encircled-count fraction of the detectors point spread function. This choice of extraction region is driven by the desire to maximize source counts, while limiting stray-light background, which is present in both \nustar\ modules in proximity to the source. Similarly, we extracted the background files from a source-free region on the same detector as the source with the same radius. For comparison purposes, we analysed two archival \nustar\ observations taken on 2016 April 23 and 2019 February 15, following the same procedures outlined above. Table~\ref{xObsLog} summarises the \nustar\ observations.

\subsection{\swift}

The X-ray Telescope (XRT) onboard \swift\ is an imaging charge-coupled device, sensitive to photons in the energy range of 0.2--10~keV \citep{burrows05SSRv}. XRT observed \src\ on several occasions in 2022 March, May and July, bracketing the time of the outburst. Thus, these observations provide the source soft X-ray spectral state shortly before the outburst onset, and a confirmation of its properties as observed with \nicer\ following the outburst. For these observations, XRT was operating in windowed-timing mode, affording a 1D image with a time resolution of 1.7~ms. We utilized {\tt XRTDAS} version 3.7.0 to reduce and calibrate the data. We extracted source events from each good time interval (GTI) of a given observation separately. We used a circular region with a 20 pixel radius centered on the brightest pixel. We extracted background events from an annulus with inner and outer radii of 80 and 120 pixels centered at the same position as the source. We created the ancillary files using \texttt{xrtmkarf}, and used the response matrices in CALDB version 20220803. We excluded all GTIs where the source landed within a 3 pixel distance from a bad column or the edge of the CCD due to increase in systematic uncertainties\footnote{\url{https://www.swift.ac.uk/user\_objects/lc\_docs.php}}. For the remaining good GTIs, we add the source and background spectra, as well as the ancillary and response files using the HEASOFT tool \texttt{addspec}. We provide the XRT observation IDs, date, and total GTI in Table~\ref{xObsLog}.

The Burst Alert Telescope (BAT), also onboard \swift, is a coded aperture telescope which images a $\sim 1.5$~sr FoV in the energy range of 15--150~keV. Images are generated by the on-board software in response to count rate increases at timescales from 4 ms to 24 s or, in the absence of rate increases, on minute and longer timescales. 
Bursts from sources outside of the FOV can often be detected due to the leakage of photons through the instrument shielding, yielding a count rate increase without a corresponding image peak.  These can be used to confirm the astrophysical nature of count rate increases seen by other instruments and constrain their direction of origin.

\subsection{\fermi-GBM}

The Gamma-ray Burst monitor \citep[GBM;][]{meegan09ApJ:gbm} is a non-imaging instrument onboard the \fermi\ Gamma-ray Space Telescope that is sensitive to photons in the 8--1000\,keV range. GBM is capable of automatic on-board triggers, with one of the software algorithms tailored to magnetar-like short bursts. Apart from the triggered data, GBM also provides continuous time-tagged events (CTTE), with a time resolution of 2 microseconds. We utilized the CTTE data to search for any untriggered bursts from the direction of \src\ from its last radio detection on 2022 March 16 until the confirmed BAT burst on 2022 April 7. This analysis provides an estimate of the start time of the 2022 outburst.

\section{Analysis and results} \label{sec:analysis}

\subsection{Radio disappearance and reactivation}

\begin{figure*}
    \centering
    \includegraphics[width=0.9\linewidth]{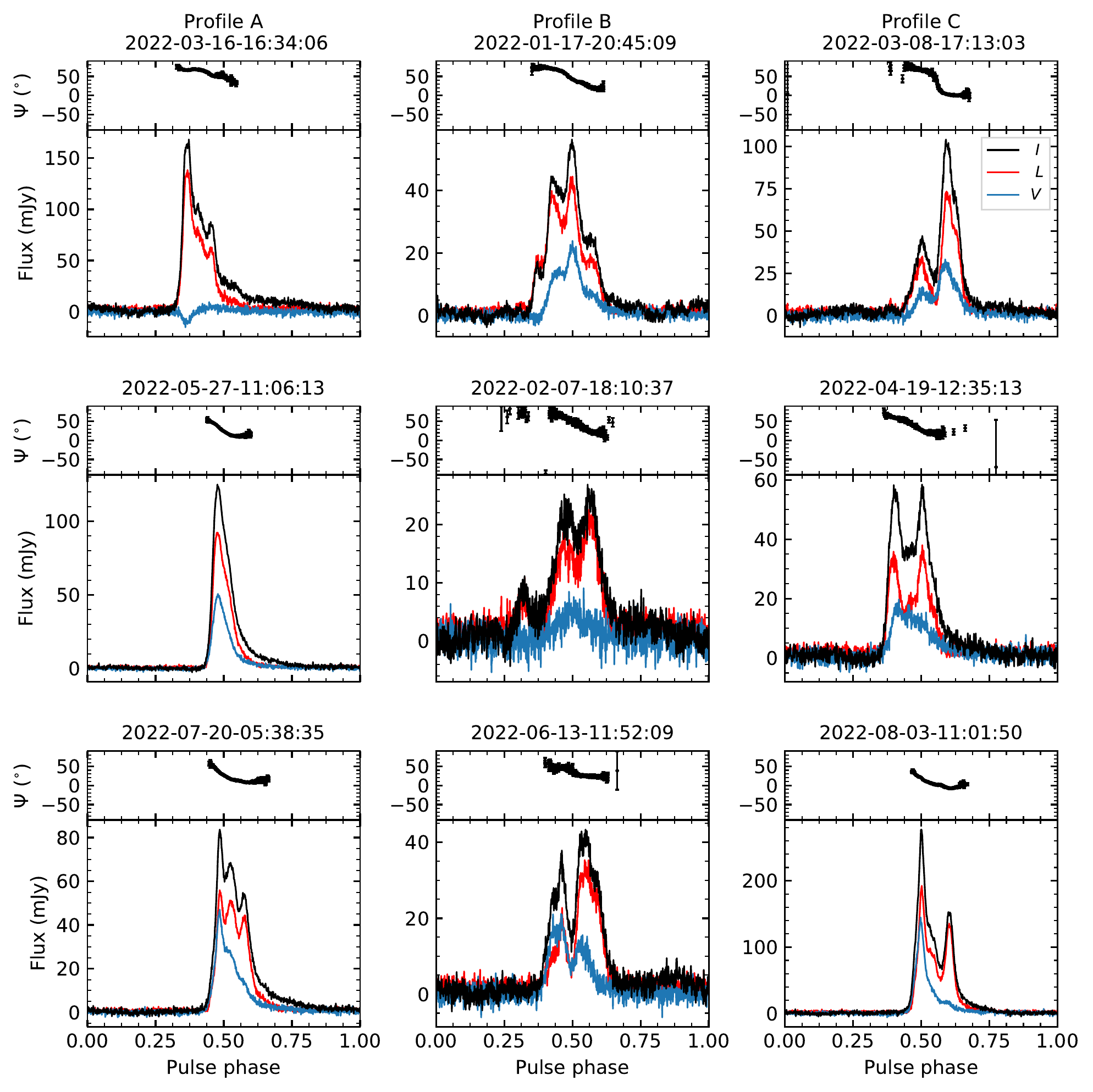}
    \caption{Examples of the three polarization profile states (A, B and C) of \src\ seen at different epochs. Note, the profiles from 2022-03-16 (top-left) and 2022-04-19 (middle-right) correspond to the last pre-outburst and first post-outburst radio detections respectively. Each observation was dedispersed using a dispersion measure of $697$\,pc\,cm$^{-3}$ and corrected for a Faraday rotation measure of $-1847$\,rad\,m$^{-2}$ prior to averaging in time and frequency.}
    \label{fig:postage_stamps}
\end{figure*}

Unlike other radio-loud magnetars which fade into radio silence in the months to years after an outburst, \src\ has remained radio-loud for well over a decade since the 2009 outburst (Camilo et al., in prep).
Following the 2008 and 2009 outbursts, detections of radio pulses were intermittent \citep[see e.g.][]{Israel2021}, possibly reflecting the highly dynamic nature of its magnetosphere during that period.
Intermittent detections continued up to 2013, after which radio pulses were consistently detected during every observation.
\src\ remained in this persistently radio-loud state until our two observations on 2022 March 16 and April 9, where we failed to detect radio pulses from the magnetar.
Radio pulses were re-detected during the 2022 April 19 observation, $\sim$12 days after the BAT confirmed trigger, and have persisted up to at least our Parkes-UWL observation on 2022 August 3.

Using a high $S/N$ observation taken on 2022 July 9, we measured a structure-optimized dispersion measure (DM) of $697 \pm 1$\,pc\,cm$^{-3}$ with the {\tt DM\_phase} package \citep{Seymour2019}.
The updated DM is a significant improvement on the previously reported value of $830 \pm 50$\,pc\,cm$^{-3}$ \citep{Camilo2007}.
This is likely due to differences in the methods applied to compute the DM (profile structure versus S/N optimisation) in addition to the substantially larger instantaneous bandwidth afforded by the UWL.
Using the NE2001 \citep{Cordes2002} and YMW16 \citep{Yao2017} electron density models, this updated DM corresponds to estimated distances of 8.3\,kpc and 5.9\,kpc, respectively. The latter distance is comparable to the commonly assumed sdistance of $\sim 4.5$\,kpc inferred from dust-scattering halos detected after the 2009 outburst \citep{Tiengo2010}, when considering the large systematic uncertainties that are associated with any of these methods, which is on average 50\%. Similarly, we obtained an updated rotation measure (RM) of $-1847.6 \pm 0.5$\,rad\,m$^{-2}$ by directly fitting the Stokes $Q$ and $U$ spectra in the 2000--4032 MHz range (to avoid complications from scatter broadening) with {\tt RMNest} \citep{Lower2022}.
Our updated RM is consistent with the previously reported value of $-1860 \pm 20$\,rad\,m$^{-2}$ \citep{Camilo2008}.

\begin{figure}
    \centering
    \includegraphics[width=\linewidth]{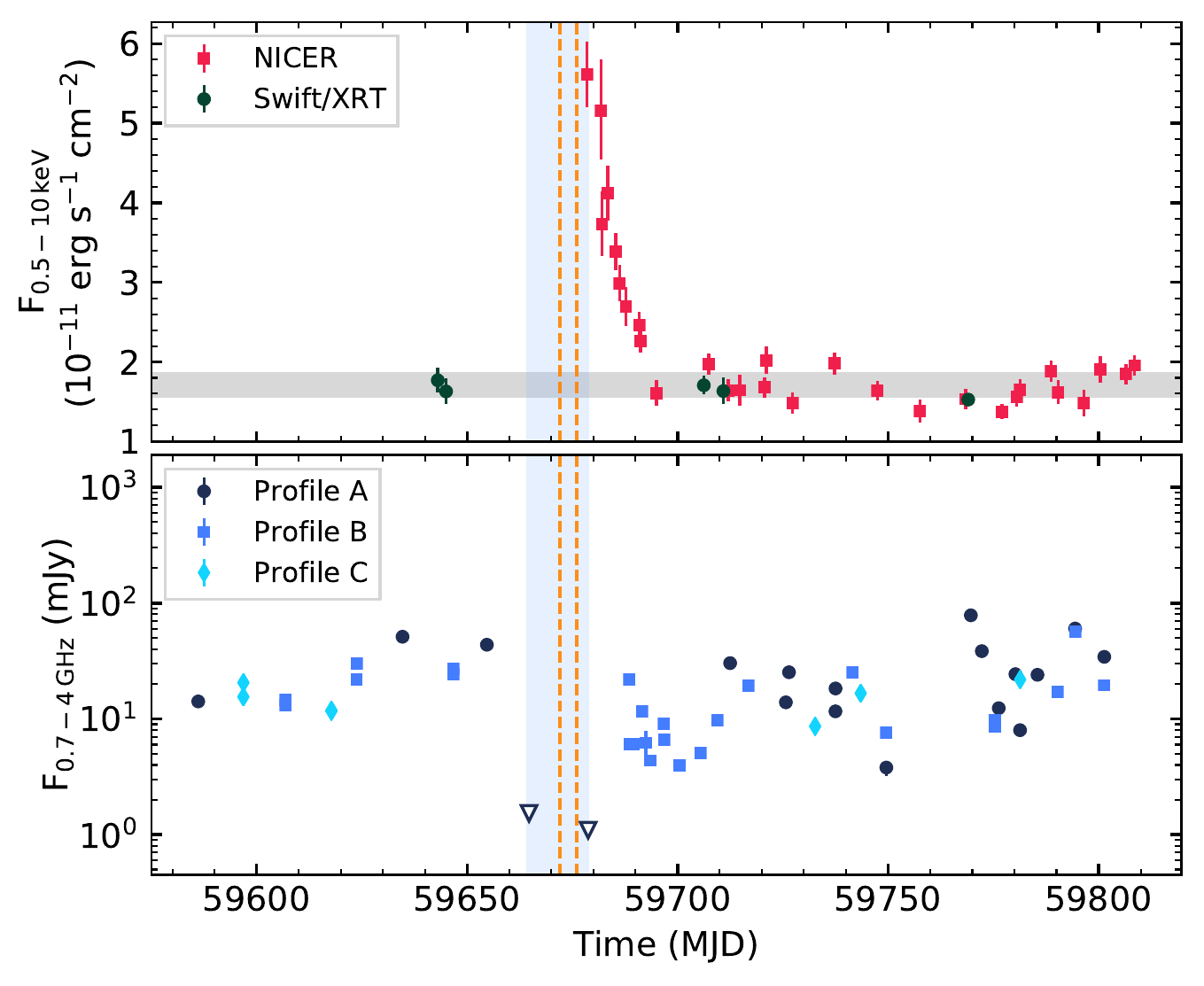}
    \caption{Average unabsorbed X-ray (0.5--10\,keV; top) and radio (0.7--4\,GHz; bottom) flux of \src\ from 2022 January to August. The vertical dashed lines indicate the dates of the \fermi-GBM (MJD 59672; 2022 April 3) and \swift-BAT short bursts (MJD 59676; 2022 April 7) respectively, while the blue shaded region corresponds to where pulsed radio emission was undetected. Grey horizontal shading corresponds to the standard deviation about the baseline X-ray flux of $1.7 \times 10^{-11}$\,erg\,s$^{-1}$\,cm$^{-2}$.}
    \label{fig:flux}
\end{figure}

The shape of the average radio profile detected throughout 2022 January-August was highly variable, and can be broadly categorized into three general archetypes:
\begin{itemize}
    \item Profile A: Strong leading component with one to three trailing components.
    \item Profile B: Complex, multi-component profile, sometimes with a central peak.
    \item Profile C: Two dominant profiles joined by weaker sub-components or a flat bridge of emission.
\end{itemize}
Examples of all three profile types are shown in Figure~\ref{fig:postage_stamps}, all of which were dedispersed and de-Faraday rotated using our updated DM and RM values.
Note, these categorizations are purely qualitative.
For instance, the complex 2022 June 13 profile (bottom-middle panel in Figure~\ref{fig:postage_stamps}) is largely double peaked, and could also be classified as Profile C.
Transitions between profile types tend to be sudden and reminiscent of transient profile changes reported in earlier observations of this magnetar (e.g., Fig. 5 of \citealt{Halpern2008}) and those seen in emission mode-changing/state-switching pulsars \citep{Wang2007}.
The radio profiles of \src\ in Figure~\ref{fig:postage_stamps} display clear epoch-to-epoch variations in flux density.
We computed the frequency-averaged radio flux density as
\begin{equation}
    {\rm F}_{f} = \frac{1}{N_{\rm bin}} \sum_{i}^{N_{\rm on}} {\rm F}_{f,i}\,,
\end{equation}
where $N_{\rm bin} = 1024$ is the number of phase-bins covering a full rotation of the magnetar, $N_{\rm on}$ is the number of bins covering the on-pulse region and ${\rm F}_{f,i}$ is the flux of the $i$-th phase bin.
Uncertainties on the flux density were computed from the root mean square (RMS) of the off-pulse region
\begin{equation}\label{eqn:flux}
    \sigma_{{\rm F},f} = \frac{\sqrt{N_{\rm on}}}{N_{\rm bin}} \sqrt{\sum_{i}^{N_{\rm off}} {\rm F}_{f,i}^{2}}\,,
\end{equation}
where $N_{\rm off}$ is the number of off-pulse phase bins.
The measured flux densities from our observations are displayed in the bottom panel of Figure~\ref{fig:flux}.
We used Equation~\ref{eqn:flux} to set sensitivity limits for our observations on 2022 March 26 and April 9 where we failed to detect radio pulses, obtaining limiting flux densities of $1.3 \pm 0.3$\,mJy and $0.9 \pm 0.2$\,mJy respectively.
Visually, there appears to be a shallow increase in the revived radio flux over time.
However, a direct connection between this apparent secular evolution and the outburst cannot be confirmed without a comparison to the long-term flux history of the magnetar.

Also indicated in Figure~\ref{fig:flux} are the corresponding profile types seen during each observation.
In terms of relative numbers, we observed the Profile B state in more than half (25/49) of our observations, while Profile A and C were seen in 17 and 7 observations respectively.
The appearance of any one mode appears to be largely stochastic, though there is a clear preference for Profile B after the outburst up until 2022 May 13 (MJD 59712)  where the magnetar was in the Profile A state.
No further systematic change in the profile or the mode-changing behavior is apparent after this date.
It is possible that this preference may be related to changes in the emission region plasma following the reactivation. 
However such profile variations persist among radio-loud magnetars for many months to years after an outburst \citep{Levin2012, Lower2021a}.
Additionally, the swing of the linear polarization position angle (PA) across the profile remains largely flat and unchanged both before and after the outburst.
It is consistent with previous observations of \src\ in 2007, where rotating vector model fits indicated that the magnetar is a near-aligned rotator \citep{Camilo2008}.
Occasional departures from the smooth PA swing are detected in individual epochs, an example of which is shown in the top-right panel of Figure~\ref{fig:postage_stamps}.
The relatively smooth evolution of the PA swing persists at higher observing frequencies where the effects of interstellar scattering are negligible, indicating these features are due to either an unusually smooth transition between two orthogonal polarization modes or the result of birefringent propagation effects in the near-field environment \citep{Dyks2020}.
Birefringent propagation effects would also explain the unusually high circular polarization fraction (up to $|V|/I = 0.54$) as well as changes in handedness between observations.
A more detailed investigation is left for future work.

\subsection{Timing}\label{timSec}

The variable radio profile of \src\ is not conducive to pulsar timing methods that use a single reference template.
Instead, we obtained pulse times of arrival (ToAs) from the \parkes/PDFB4 data by using the {\tt get\_toas} tool in {\tt PRESTO} \citep{Ransom2011} to fit a single Gaussian template to individual observations via the Fast Fourier Transform based approach of \citet{Taylor1992}.
We added an extra 0.05\,s uncertainty in quadrature with the formal ToA uncertainties to account for excess epoch-to-epoch scatter in the ToAs that arise from the varying radio profile (often referred to as `jitter').

For \nicer, we merged all barycenter-corrected event files and only accepted counts in the 1--5~keV energy range, which maximized the source pulse signal according to the $Z^2$ power \citep{buccheri83AA}. We then split the data into time intervals that contain $\approx1000$~counts each, without allowing individual intervals to exceed a 3-day integration. We employed a non-binned maximum likelihood estimation (MLE) method to measure the X-ray pulse ToAs \citep{livingstone09ApJ,ray11ApJS}. Firstly, we model a high S/N pulse profile with a Fourier series including only the fundamental harmonic; adding higher order harmonics did not improve the quality of the fit. Secondly, we fit, using MLE, each unbinned data segment to the same model, only allowing for a phase-shift $\Delta\phi$. The best-fit phase shift is the value that corresponds to the minimum of the negative log likelihood, $-\log(\mathcal{L})_{\rm min}$. The $1\sigma$ uncertainty on the phase shift was established by stepping over the full parameter space $\Delta\phi \in [0,2\pi)$ and noting a change in $-\log(\mathcal{L})_{\rm min}$ by $0.5$ (corresponding to the $\pm34\%$ or $1\sigma$ confidence interval of a $\chi^2$ distribution with 1~degree of freedom).

We obtained an initial phase-connected solution by using the Hidden Markov Model (HMM) based glitch detector of \citet{Melatos2020}\footnote{\href{https://github.com/ldunn/glitch_hmm}{https://github.com/ldunn/glitch\_hmm}} which tracks the spin-frequency ($\nu$) and spin-down rate ($\dot{\nu}$) of the magnetar over time. 
This enabled robust integer pulse numbering to be applied to each ToA and provided estimates of any changes in $\nu$ and $\dot{\nu}$ at the time of the outburst.

\begin{figure}
    \centering
    \includegraphics[width=\linewidth]{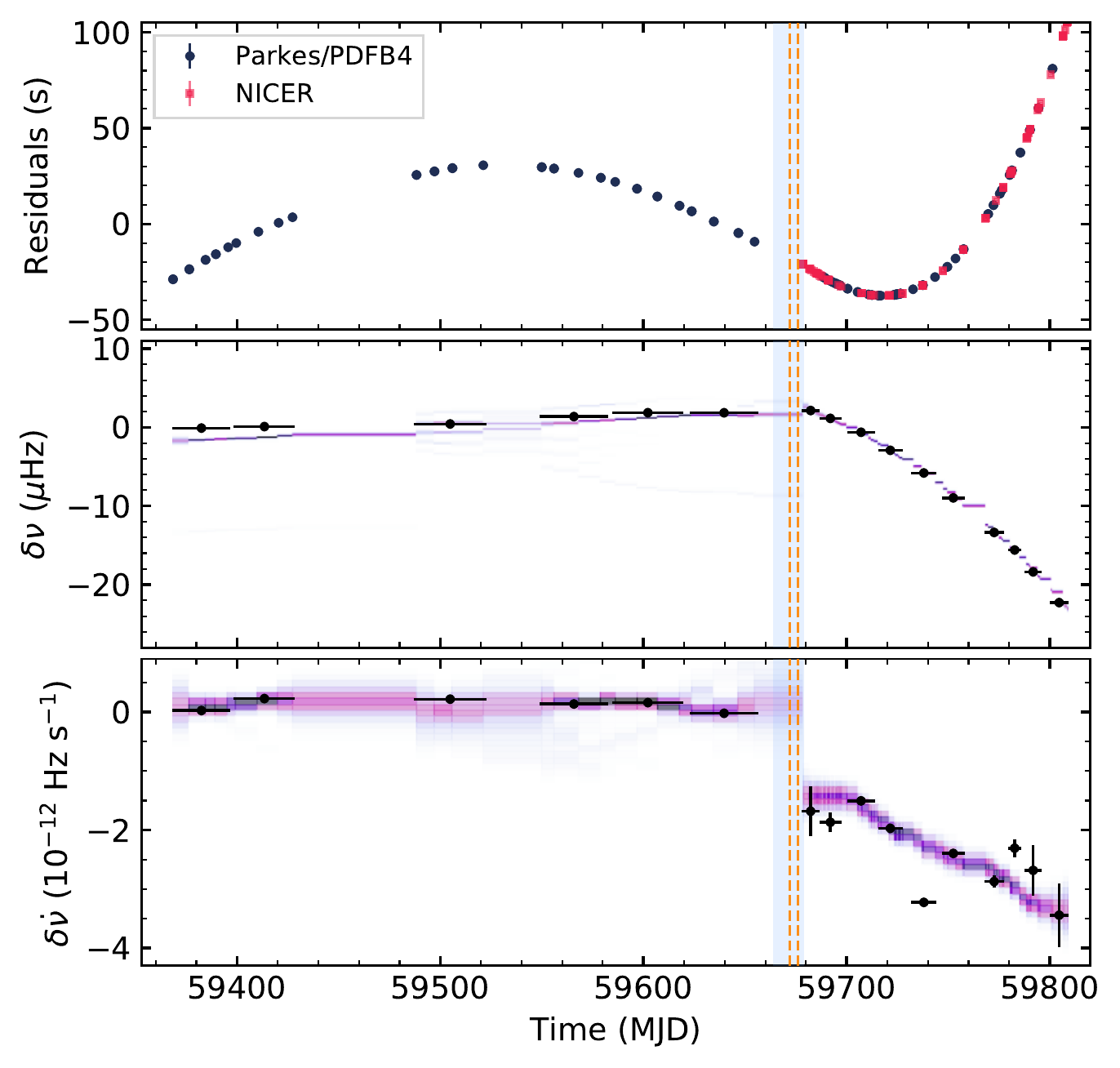}
    \caption{Top panel shows the pre-fit \parkes/\nicer\ timing residuals of \src\ from 2021 June to 2022 August. Changes in the magnetar spin-frequency and spin-down rate with respect to their nominal values ($0.478836$\,Hz and $-3.63 \times 10^{-12}$\,s$^{-2}$ at MJD 58350) are shown in the middle and bottom panels respectively, where the points correspond to local fits to the ToAs. Purple shading is the $\nu$ and $\dot{\nu}$ posterior distributions that were recovered by the HMM glitch detector at each epoch. Outliers in the local fits to $\dot{\nu}$ arise from the limited number of ToAs. Vertical shaded region and dashed lines are the same as in Figure~\ref{fig:flux}.}
    \label{fig:f0f1_fits}
\end{figure}

Once the phase-connected solution was obtained, we then used the {\tt glitch} plugin to {\tt tempo2} \citep{Hobbs2006, Edwards2006} to perform independent local fits to $\nu$ and $\dot{\nu}$.
The results of our local fits are shown alongside the HMM tracking in Figure~\ref{fig:f0f1_fits}.
Both sets of spin-frequency/spin-down rate measurements generally agree with one another and display approximately the same secular evolution.
A timing anomaly with clear jumps in both $\nu$ and $\dot{\nu}$ occurred in coincidence with the high-energy outburst, with the HMM recovering step-changes of $\Delta\nu = (9 \pm 4) \times 10^{-7}$\,Hz and $\Delta\dot{\nu} = (-1.6 \pm 0.5) \times 10^{-12}$\,s$^{-2}$. 
There is also a clear, secular increase in spin-down rate of the magnetar ($\dot{\nu}$ is becoming more negative) over time, where $\dot{\nu}$ appears to follow an approximately linear trend over the first 120\,days post-outburst.
This is also reflected by the downward curve in the evolution of $\nu$ over time.
Increasing spin-down rates are also commonly observed in the post-outburst behavior of other magnetars, including recent outbursts in the radio-loud magnetars PSR~J1622$-$4950 \citep{Camilo2018} and XTE~J1810$-$197 \citep{Caleb2022}.
A linear fit to the spin-down evolution gives an apparent $\ddot{\nu} = -1.95(5) \times 10^{-19}$\,s$^{-3}$.
Similar behavior was previously observed in \src\ following the 2008 outburst \citep{Israel2010}.

In order to better constrain both the longer-term rotational properties of \src\ and the outburst-induced changes in its timing, we modelled the anomaly as a pulsar glitch where the corresponding change in rotation phase is given by
\begin{eqnarray}
     \phi_{\rm {g}}(t) = \Delta\phi_{g} + \Delta\nu_{p}(t - t_{\rm {g}}) + \frac{1}{2}\Delta\dot{\nu}_{p}(t - t_{\rm {g}})^{2} \nonumber \\
     + \frac{1}{6}\Delta\ddot{\nu}_{p}(t - t_{\rm {g}})^{3} - \Delta\nu_{d}\tau_{d} e^{-(t - t_{\rm {g}})/\tau_{d}}.
\end{eqnarray}
Here, $t_{g}$ is the glitch epoch, $\Delta\phi_{g}$ is a phase-jump to account for residual ambiguities in the rotation phase of the magnetar after the event, $\Delta\nu_{p}$ is the permanent step-change in spin-frequency, $\Delta\dot{\nu}_{p}$ the step-change in spin-down rate, and $\Delta\nu_{d}$ is the exponentially recovered change in spin-frequency that occurred over a characteristic timescale, $\tau_{d}$.
We also included a glitch-induced second spin-frequency derivative ($\Delta\ddot{\nu}_{p}$) term to account for the post-outburst linear increase in $\dot{\nu}$ that we discussed above.

The gap in our timing between the last \parkes\ detection on 2022 March 16 and the first \nicer\ observation on April 9 makes it difficult to determine the precise glitch epoch. Since magnetar glitches are often associated with radiative events \citep{dib14ApJ}, we use the date of the short-burst detected by \fermi-GBM from the direction of \src\ as the nominal glitch epoch (i.e $t_{g} = $\,MJD 59672; 2022 April 3; see Sec.~\ref{sec:burSear}). 
Note, the exact choice of glitch epoch between April 1 and April 7 has a strong impact on the recovered $\Delta\nu_{p}$, which we discuss further below. 
Using {\tt tempo2} and the {\tt FITWAVES} plugin, we fit the glitch parameters alongside a set of 10 harmonically related sine-waves to account for strong timing noise in the residuals. 
This returned values of $\Delta\nu_{p} = 2.0(1) \times 10^{-6}$\,Hz, $\Delta\dot{\nu}_{p} = -1.23(2) \times 10^{-14}$\,s$^{-2}$ and $\Delta\ddot{\nu}_{p} = -1.86(3) \times 10^{-19}$\,s$^{-3}$.
A significant $\Delta\nu_{d}$ of $-2.1(1) \times 10^{-6}$\,Hz was recovered. The negative value of $\Delta\nu_{d}$ signifies a resolved rise in the spin-frequency over a period of $\tau_{d} = 13(1)$\,days as opposed to a post-glitch decay.

To verify the existence of this apparent delayed rise in $\nu$, we sampled the posterior distributions for each parameter using the {\tt TempoNest} Bayesian inference plugin to {\tt tempo2} \citep{Lentati2014}.
This also allowed us to simultaneously model the timing noise of the magnetar as a power-law red noise process. The power-spectral density of the red noise model is given by
\begin{equation}
    P(f) = \frac{A_{\rm r}^{2}}{12\pi^{2}} \Big( \frac{f}{1\,{\rm yr}} \Big)^{-\beta_{\rm r}}\,{\rm yr}^{-3},
\end{equation}
where $A_{\rm r}$ is the red noise amplitude (in units of yr$^{3/2}$) and $\beta_{\rm r}$ the spectral index.
Two separate fits were performed: one that included a resolved rise in spin frequency, and another where $\Delta\nu_{d}$ is fixed to zero for an instantaneous (unresolved) glitch.
Comparing the recovered log-evidence from the resulting fits, we obtain a Bayes' factor of $\ln\mathcal{B} = -2.8$ in favor of a resolved rise.
The negative value indicates there is no significant evidence for the presence of a resolved versus an instantaneous change in spin-frequency.
Our results for the un-resolved glitch model are summarized in Table~\ref{tab:timing}.

\begin{table}[t!]
\centering 
\caption{Timing and pulsar parameters for \src\ from our joint \parkes/\nicer\ timing. Values in parentheses are the 1-$\sigma$ uncertainties on the last digit.}
\label{tab:timing}
\begin{tabular}{ll}
\hline
\hline
Pulsar parameters &  \\
\hline
R. A. (J2000)$^{\dagger}$                    & 15$^{\mathrm{h}}$50$^{\mathrm{m}}$54.12386$^{\mathrm{s}}$ \\
Decl. (J2000)$^{\dagger}$                    & $-$54$^{\circ}$18$'$24.1141$''$ \\
$\nu$ (Hz)                                   & $0.478836(3)$ \\
$\dot{\nu}$ (s$^{-2}$)                       & $-3.63(3) \times 10^{-12}$ \\
$\mu_{\alpha} \cos\delta$ (mas)$^{\dagger}$  & $4.8$ \\
$\mu_{\delta}$ (mas)$^{\dagger}$             & $-7.9$ \\
Timing epoch (MJD TDB)                       & 58350 \\
Position epoch (MJD TDB)                     & 54795 \\
DM (pc\,cm$^{-3}$)                           & $697(1)$ \\
RM (rad\,m$^{-2}$)                           & $-1847.6(5)$ \\
Time span (MJD)                              & 59368--59808 \\
Number of ToAs                               & 202 \\
Solar system ephemeris                       & DE436 \\
$\log_{10}(A_{\rm r})$ (yr$^{3/2}$)          & $-5.1(2)$ \\
$\beta_{\rm r}$                              & $5.7(8)$ \\
\hline
Glitch parameters &  \\
\hline
$t_{\rm g}$ (MJD)                   & 59672 \\
$\Delta\nu_{p}$ (Hz)                & $2(1) \times 10^{-7}$ \\
$\Delta\dot{\nu}_{p}$ (s$^{-2}$)    & $-2.4(1) \times 10^{-12}$ \\
$\Delta\ddot{\nu}_{p}$ (s$^{-3}$)   & $-2.0(1) \times 10^{-19}$ \\
\hline\\
\end{tabular}
{$^{\dagger}$}Fixed to values from~\cite{Deller2012}.
\end{table}

\begin{figure}[t!]
\begin{center}
\includegraphics[width=\linewidth]{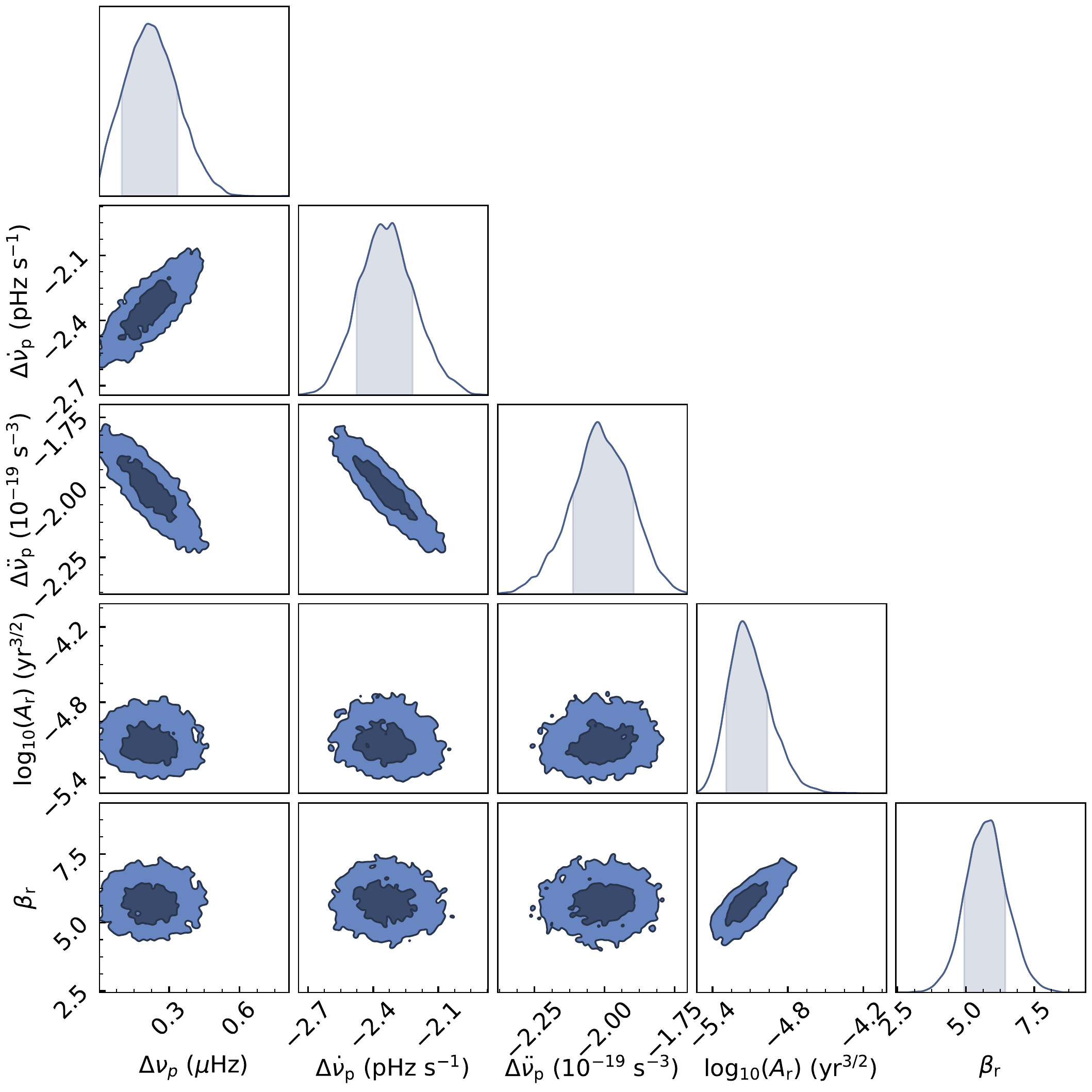}
\caption{One- and two-dimensional posterior distributions for the glitch and red noise in \src. Darker and lighter shaded regions represent the 68\% and 95\% confidence intervals respectively.}
\label{fig:glitch_post}
\end{center}
\end{figure}

We present the one- and two-dimensional posterior distributions for the glitch and timing noise parameters in Figure~\ref{fig:glitch_post}. There is a strong correlation between $\Delta\dot{\nu}_{p}$ and $\Delta\ddot{\nu}_{p}$, indicating these two parameters induce a similar signal across the relatively short post-glitch timing of \src. While $\Delta\dot{\nu}_{p}$ agrees with the value recovered by the HMM at the 68\% confidence interval, the {\tt TempoNest} derived value of $\Delta\nu_{p}$ is smaller by a factor of $\sim$4.5. As for the post-outburst rotational-evolution, our recovered value of $\Delta\ddot{\nu}_{p}$ is consistent with that inferred from a simple linear fit to $\dot{\nu}$ over time.

The difference in $\Delta\nu_{p}$ values is a direct result of our choice of glitch epoch and the large increase in spin-down rate. This is apparent in {\tt TempoNest} runs where we assumed earlier and later values of $t_{\rm g}$, where earlier glitch epochs result in a larger recovered value of $\Delta\nu_{p}$ to match the post-outburst timing. Going the other way, setting the glitch epoch to the date of the \swift-BAT short burst (MJD 59676; 2022 April 7) results in a value of $\Delta\nu_{p}$ that is consistent with zero at the 68\% confidence interval, i.e a pure spin-down glitch. While several glitches in other magnetars have been associated with minimal instantaneous changes in $\nu$ (e.g. the 2012 outburst of 1E~1048.01$-$5937; \citealt{Archibald2015}), the lack of ToAs between the radio shutdown of \src\ and beginning of our \nicer\ monitoring prevents us from ruling out a non-zero $\Delta\nu_{p}$.

\subsection{High-energy outburst}

\subsubsection{Spectral analysis}

We employed Xspec version 12.12.1 for our X-ray spectral analysis. We used the \texttt{tbabs} model to account for the line-of-sight absorption towards the source, along with the \citet{wilms00ApJ} solar abundances. We binned all spectra to have 5 counts per bin and used the W-statistic (C-stat in Xspec) to estimate the best fit model parameters and their associated uncertainties. For the \nicer\ spectra, however, we used the pgstat statistic, which is a more appropriate choice given the modeled nature of the background \citep{remillard2021:3c50}. Finally, we added a numerical factor factor to account for any cross-calibration uncertainties when fitting spectra from different instruments simultaneously. We find a $\lesssim2\%$ and $\lesssim10\%$ deviation between the two \nustar\ spectra and \nicer/\swift-XRT and \nustar\ spectra, respectively, commensurate with the expected values \citep{madsen15ApJS}.

\begin{figure}[t!]
\begin{center}
\hspace{-0.0cm}
\includegraphics[angle=0,width=0.48\textwidth]{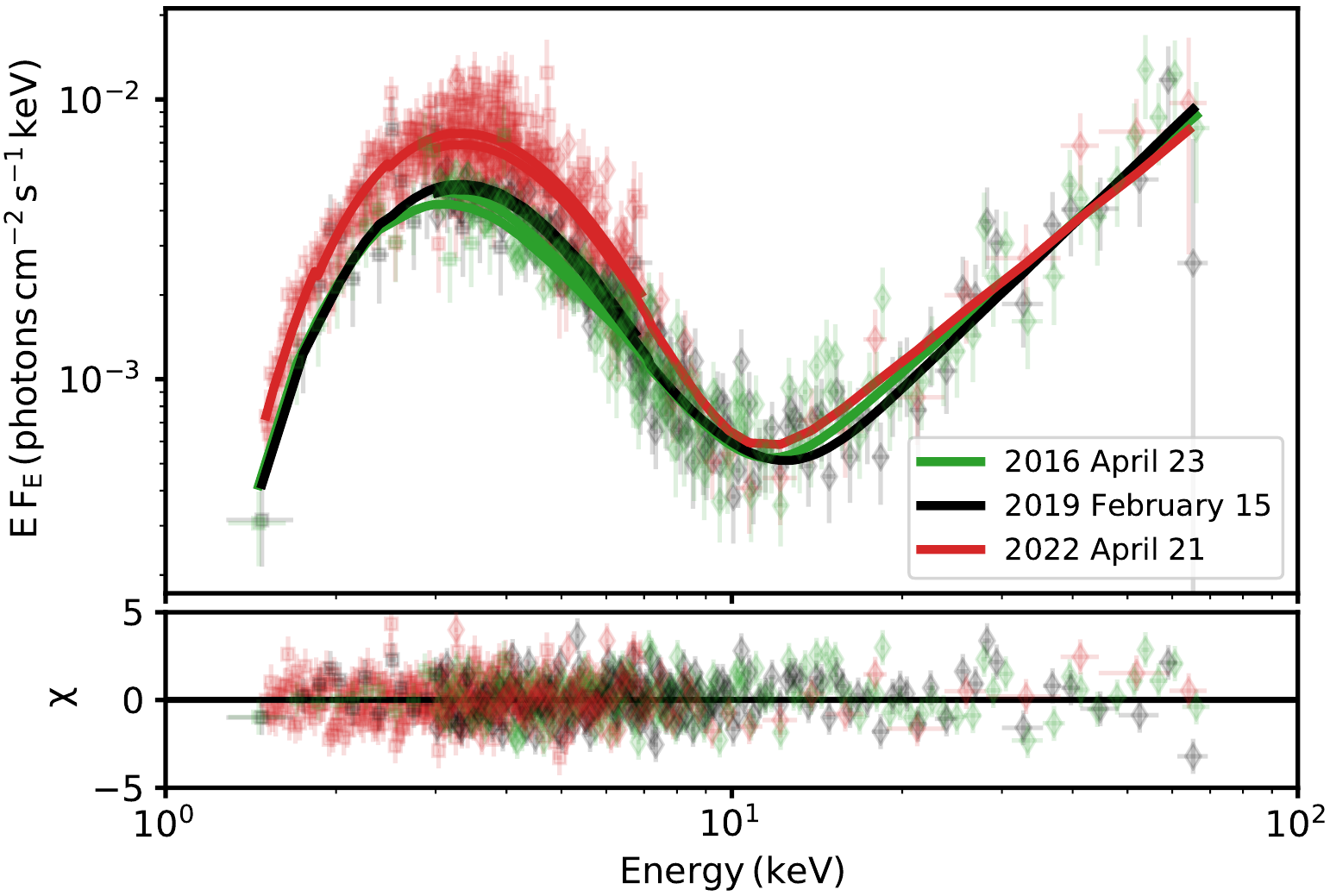}
\caption{{\sl Upper panel.} Broad-band $\rm \nu F_\nu$ spectra of \src\ derived during the 2022 April outburst (\nicer+\nustar, red), 2019 February (XRT+\nustar, black) and 2016 April (XRT+\nustar, green). Solid lines represent the best-fit model consisting of two BB and a PL. A flux enhancement is clearly present at soft X-rays coincident with the most recent outburst, in contrast to the stable non-thermal hard X-ray tail. {\sl Lower panel.} Deviation of the data from the best-fit model in units of $\sigma$.}
\label{specFig}
\end{center}
\end{figure}

We fit the \nustar\ and \nicer\ observations taken on 2022 April 21 simultaneously, which provides the highest S/N broad-band spectrum shape of \src\ during this most recent outburst. We also fit the pre-outburst 2016 and 2019 simultaneous \swift-XRT and \nustar\ observations \citep{CotiZelati2020} to better quantify the source spectral variability. We modelled all spectra in the 1--70~keV range using a three-component construction: two absorbed blackbodies ($\rm BB_{warm}$ and $\rm BB_{hot}$) plus a power-law (PL), suitably representative of the magnetar broad-band X-ray spectral shape \citep{Kuiper-2006-ApJ,Goetz-2006-AandA,denHartog-2008a-AandA,denHartog-2008b-AandA,enoto17ApJS}. We linked the absorption hydrogen column density $N_{\rm H}$ among all the spectra and left the rest of the model parameters free to vary. We establish the goodness of fit by running the Xspec \texttt{goodness} command to simulate $1000$ spectra drawn from the best-fit model and compare their fit statistic to the observed data. We find that 16\% of the spectra have a fit statistic smaller than the actual value, implying that the model adequately explains the data. We measure a best-fit $N_{\rm H} = (3.67 \pm 0.02) \times 10^{22}$~cm$^{-2}$, a factor of 1.75 times higher than the value predicted by the $N_{\rm H}$-DM relation of \citet{He2013}. We display the source spectrum and best fit model in Figure~\ref{specFig}, and present the best fit parameters and their $1\sigma$ uncertainty in Table~\ref{specParam}.

\begin{table*}[t]
\caption{Broad-band X-ray spectral results}
\label{specParam}
\newcommand\T{\rule{0pt}{2.6ex}}
\newcommand\B{\rule[-1.2ex]{0pt}{0pt}}
\begin{center}{
\hspace{-2.0cm}
\resizebox{1\textwidth}{!}{
\begin{tabular}{l c c c c c c c c}
\hline
\hline
Date & $kT_{\rm warm}$ \T\B & $R^2_{\rm warm}$ & $kT_{\rm hot}$ & $R^2_{\rm hot}$ & $\Gamma$ & $F_{\rm warm}$/$F_{\rm hot}$/$F_{\rm PL}$ & $F_{\rm S}$/$F_{\rm H}$ & $L_{\rm X}$\\
& (keV)       \T\B     &    (km$^2$)          &  (keV)            & (km$^2$)          &                     & ($10^{-12}$~erg s$^{-1}$ cm$^{-2}$)                         &   ($10^{-12}$~erg s$^{-1}$ cm$^{-2}$) &  ($10^{34}$~erg s$^{-1}$) \\
\hline
2016-03-23 & $0.59\pm0.03$ \T\B & $2.3_{-0.5}^{+0.7}$ & $1.16_{-0.08}^{+0.1}$ & $0.03_{-0.01}^{+0.02}$ & $0.2\pm0.1$ & $14.1_{-0.8}^{+0.9}$ / $2.9_{-0.6}^{+0.7}$ / $8.9_{-0.5}^{+0.5}$ & $17.2_{-0.8}^{+0.9}$ / $8.7_{-0.4}^{+0.4}$ & $6.3_{-0.2}^{+0.2}$ \\
2019-02-15 & $0.67\pm0.02$ \T\B & $1.3\pm0.2$       & $1.4\pm0.2$ & $0.008\pm0.004$ & $0.0\pm0.1$ & $13.7_{-0.5}^{+0.5}$ / $1.7_{-0.3}^{+0.5}$ / $8.7_{-0.6}^{+0.7}$ & $15.5_{-0.5}^{+0.5}$ / $8.7_{-0.5}^{+0.5}$ & $5.9_{-0.1}^{+0.2}$ \\
2022-03-21 & $0.64_{-0.05}^{+0.04}$ \T\B & $2.0_{-0.3}^{+0.5}$ & $1.0\pm0.1$ & $0.09_{-0.06}^{+0.14}$ & $0.4\pm0.2$ & $17_{-3}^{+2}$ / $6_{-2}^{+3}$ / $9.0_{-1}^{+1}$ & $23_{-1}^{+1}$ / $8.7_{-0.8}^{+0.8}$ & $7.7_{-0.3}^{+0.3}$ \\
\hline
\hline
\end{tabular}}}
\end{center}
\begin{list}{}{}
\item {\bf Notes.} BB emitting areas and luminosity are derived assuming a distance of 4.5~kpc \citep{Tiengo2010}. $F_{\rm S}$ and $F_{\rm H}$ are the 0.5-10 keV and 10-70 keV absorption-corrected fluxes, respectively.
\end{list}
\end{table*}

We find that the 2022 April 21 observation, 2.5 weeks after the GBM short burst detection (2 weeks after the confirmed BAT detection), exhibits a flux that is about 50\% larger than the pre-outburst level. This increase is solely due to an increase in the soft 0.5--10~keV X-ray flux (Figure~\ref{specFig}), and particularly seen in the $\rm BB_{hot}$ contribution. The flux of this component has increased by a factor 3 and 5 since 2016 and 2019, respectively. On the other hand, the increase in the $\rm BB_{warm}$ component is a factor 1.2  and 1.3 larger compared to 2016 and 2019, respectively. Finally, the hard X-ray PL flux is consistent between all three observations at the $1\sigma$ level. This implies that either the hard X-ray flux was not impacted by this most recent outburst, or it followed a steeper temporal decay trend compared to the soft X-ray flux.

\begin{figure}[h!]
\begin{center}
\includegraphics[angle=0,width=0.48\textwidth]{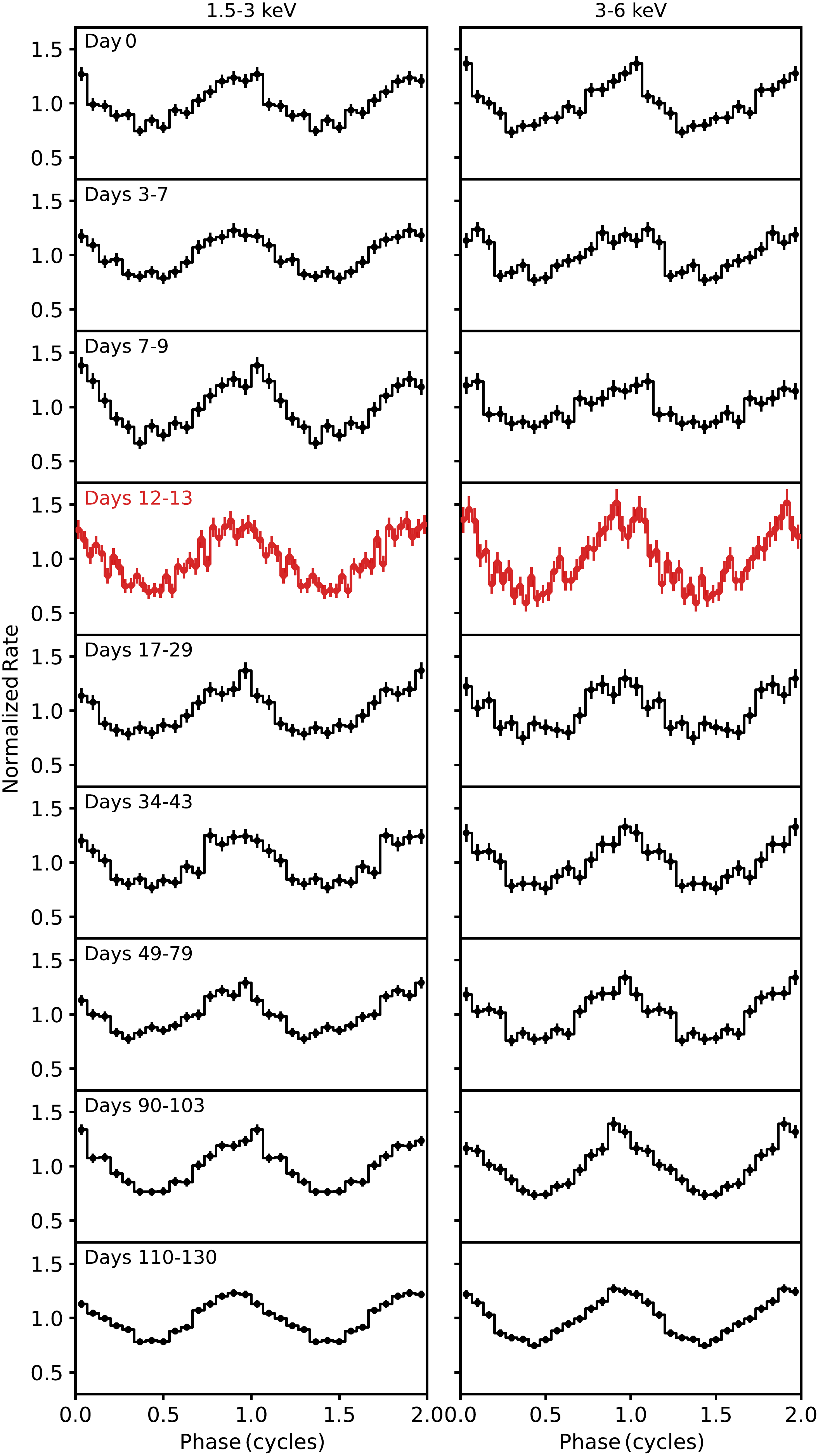}
\caption{Time-resolved \nicer\ pulse profiles of \src\ in days from 2022 April 9. We show the pulsations in the energy range 1.5--3~keV (left column) and 3--6~keV (right column). The red profile corresponds to the \nicer\ observation simultaneous to \nustar\ taken on 2022 April 21.  All profiles are displayed using 15 phase bins per rotational cycle, with the exception of the red profile, which exhibits 30 phase bins per pulse period (see text for details). Two cycles are shown for clarity.}
\label{nicProf}
\end{center}
\end{figure}

While the source exhibited a change in the overall flux, the broad-band spectral curvature in 2022 remained relatively stable compared to the previous epochs. We find no significant variability in the temperature of either of the two BBs, nor in the hardness of the PL component. This points to an increase in the overall area of the hot thermally-emitting region as the origin of the soft X-ray flux increase (Table~\ref{specParam}). 

To study the long-term high energy evolution of the \src\ outburst, we modelled the \nicer\ and \swift-XRT spectra in the 1--6~keV range (beyond 6~keV the data for both instruments was mainly background-dominated). We used a simple absorbed BB model and fixed the hydrogen column density $N_{\rm H}$ to the best-fit value as derived above. This model sufficed to describe all the spectra except for the \nicer\ observation on 2022 April 9 (ID 5020300101, Table~\ref{xObsLog}) which required a second blackbody component to result in a statistically acceptable fit. The resulting 0.5--10~keV absorption corrected flux evolution is shown in the upper-panel of Figure~\ref{fig:flux}. At the start of the \nicer\ monitoring on 2022 April 9, we find a source flux that is a factor 3.5 larger than the pre-outburst value as established with \swift-XRT on 2022 March 5 (green circles in Figure~\ref{fig:flux}, upper panel; see also \citealt{CotiZelati2020}). This confirmed that \src\ was experiencing yet another high energy outburst, its first since its major 2009 outburst. The source flux decayed back to the same pre-outburst persistent flux level on 2022 April 26, i.e. 17 days later, and remained there until the end of our monitoring in 2022 August. We find that the bulk of the flux increase throughout the outburst occurred at energies $>2$~keV, consistent with the result of the \nicer+\nustar\ broad-band modelling.

\subsubsection{X-ray timing analysis}

Given the complexity of the glitch timing model  (Section~\ref{timSec}), we fold the \nicer\ barycenter-corrected events with a (strictly) post-glitch variant of the combined radio/X-ray timing solution in Table~\ref{tab:timing}.
To account for red noise in the timing residuals, we fit for spin-frequency derivatives up to $d^5\nu/dt^5$, at which order the residuals are dominated by white noise. Due to the limited exposure of most individual \nicer\ observations, we merged several consecutive ones until we accumulated about 6000 counts in the 1.5--6~keV interval. This ensured a proper sampling of the pulse shape in two energy bands, namely 1.5--3~keV and 3--6~keV. During the outburst decay phase (2022 April 9-26), the stretch of each interval that fulfilled our criterion was on average 2 days, and increased to 10~days after the source returned to its pre-outburst flux level. We present these pulse profiles in Figure~\ref{nicProf}.  Each row displays the pulsations per time interval starting from the date of the initial \nicer\ observation on 2022 April 9 (day 0), while the left and right panels show the 1.5--3~keV and 3--6~keV pulse-profiles, respectively; note that the pulse in red is from the \nicer\ observation that is simultaneous to the \nustar\ one for 2022 April 21.

The soft 1.5--3~keV pulsations are quasi-sinusoidal throughout our monitoring campaign. On the other hand, the 3--6 keV profile exhibits pronounced variability. For instance, during the outburst decay phase, there is an evident shoulder leading to the pulse peak. Moreover, the peak is slightly broader early in the outburst and displays complex shape, likely formed of two or more peaks. This complexity is confirmed with \nustar\ as we show below. The 3--6 keV pulse shape turns quasi-sinusoidal starting on day 17 from the first \nicer\ observation, coincident with the return of the flux to the pre-outburst level (Figure~\ref{fig:flux}). We derive the root-mean-square pulsed fraction (rms PF) for each of the light curves in Figure~\ref{nicProf} \citep[see, e.g.,][for details]{woods04ApJ}. For this purpose, we fit each background-subtracted pulse profile to a Fourier series including three harmonics, which was sufficient to provide an acceptable fit to the data. We find pulsed fractions ranging from 19--25\% and 29--34\% in the 1.5--3~keV and 3--6~keV energy bands, respectively, with a typical $1\sigma$ uncertainty of about 3\%. This implies a lack of significant variability in the pulsed fraction at soft X-rays concurrent with the outburst flux enhancement.

\begin{figure}[t!]
\begin{center}
\includegraphics[angle=0,width=0.48\textwidth]{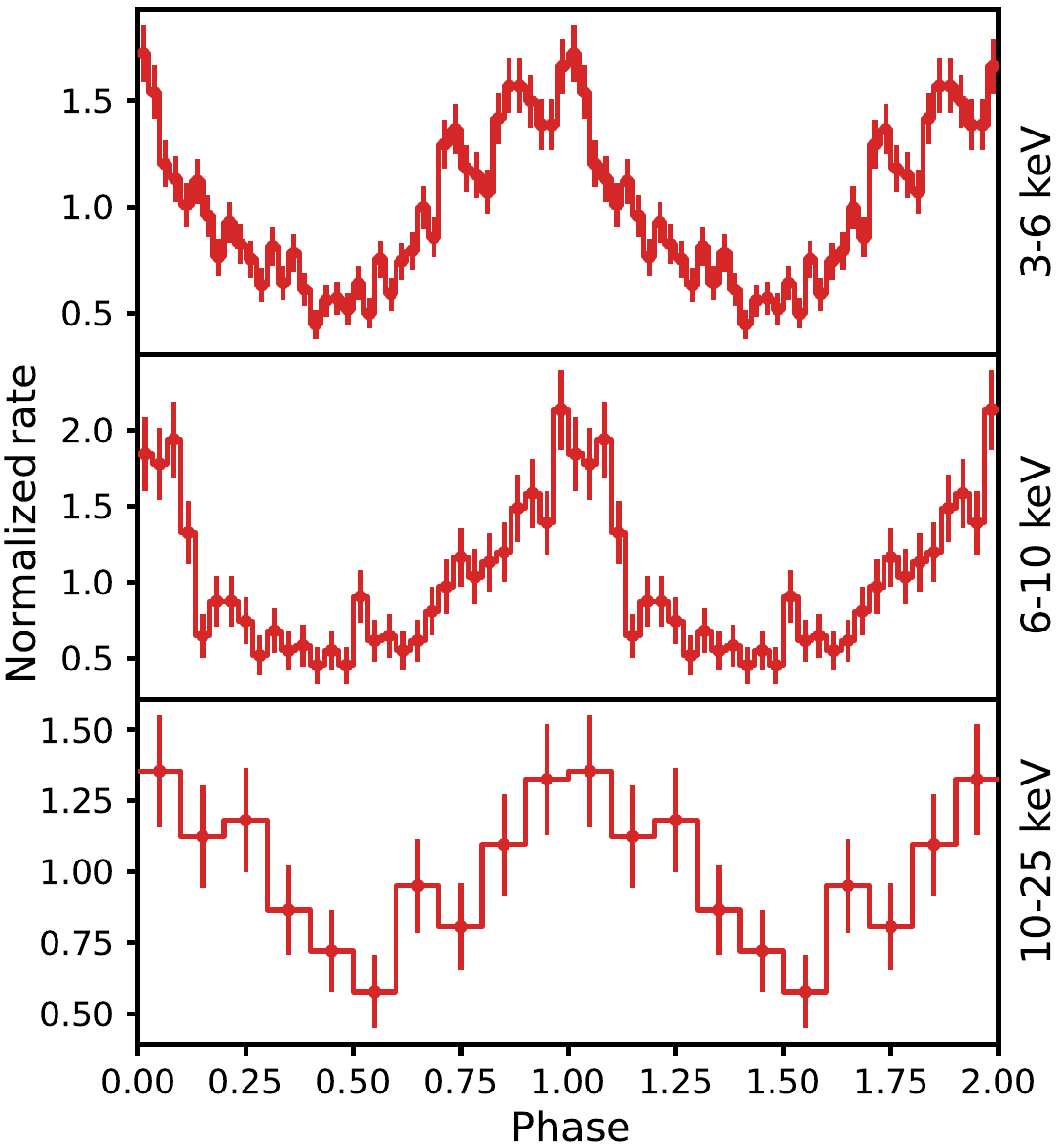}
\caption{The \src\ pulse profiles as derived with \nustar\ at different energies (top to bottom). See text for more details.}
\label{Xprof}
\end{center}
\end{figure}

The \nustar\ observation of \src\ on 2022 April 21 caught the tail end of the outburst. We display the pulse profiles during this observation in 3 separate energy bands, namely, 3--6~keV, 6--10~keV, and 10--15~keV (Figure~\ref{Xprof}). We confirm the \nicer\ results which show, in the energy range 3--6 keV, a non-symmetric profile with a shoulder leading to a multi-peaked pulse maximum; this fine structure in the peak is reminiscent of that seen during the 2020 outburst of SGR 1830-0645 \citep{younes22ApJL18301}.  Similar pulse profile morphology is apparent in the 6--10 keV range: the pulse exhibits considerable asymmetry with a shoulder leading to the pulse peak. All these features are clearly absent from the 2016 and 2019 \nustar\ monitoring; the pulse profile in the same energy range shows a purely sinusoidal form \citep[][Figure~5]{CotiZelati2020}. On the other hand, the pulse profile at high energies between 10 and 25~keV exhibits quasi-sinusoidal shape, consistent with the result of the earlier \nustar\ observations. Note, however, the lower S/N in this band compared with that at energies below 10 keV. We do not detect significant pulsations at energies $>25$~keV, despite the detection of the source at the $\gtrsim10\sigma$ confidence level.

Finally, we measure the \nustar\ energy-resolved rms PF following the same recipe we utilized for \nicer. We find an increasing trend at soft energies, from about $30\%$ and up to $45\%$ at 10 keV. The PF drops to $30\%$ in the energy range 10--25~keV. Moreover, the latter profile appears slightly offset from the soft-energy ones (Figure~\ref{Xprof}), yet a formal measurement is complicated by the difference in shape between the profiles. Lastly, we derive a 3$\sigma$ upper-limit of $25\%$ on the pulsed fraction at energies $>25$~keV. These elements are broadly consistent with the 2016 and 2019 \nustar\ findings of \citet{CotiZelati2020}.

\subsubsection{Short burst searches}
\label{sec:burSear}

The 22-day discrepancy between the time of the first radio non-detection, which occurred on 2022 March 16, and the BAT trigger on 2022 April 7 prompts the question on whether the radio disappearance genuinely preceded the trigger of the high energy outburst. With the absence of any sensitive X-ray observations of \src\ during this gap, we relied on burst detection to qualitatively gauge the source activity. Typically, the rate of hard X-ray bursts from an active magnetar is maximal at the start of an outburst, declining thereafter. This behavior is observed during burst storms, e.g., 1E~2259+596 \citep{gavriil04ApJ2259}, SGR 1935+2154 \citep{Younes2020}, as well as over the course of weeks following the magnetar outburst onset, e.g., SGR 1830$-$0645 \citep{younes22ApJ18301}, Swift~J1818.0$-$1607 \citep{hu20ApJ}. Hence, if the outburst indeed began at the time of the radio non-detection, more bursts should be detected from the direction of \src\ by Fermi-GBM during the 20~day interval.

We searched the CTTE data for untriggered bursts in the energy range 8--100~keV following the procedure detailed in \citet{Gavriil2004} and \citet{Younes2020}. In summary, we binned the light curve of each GBM detector separately, at a temporal resolution $t_{\rm r}$. We search each light curve in consecutive intervals of 100~s, overlapping by 10~s. We note the deviation of the number of counts $x$ in each bin from the long-term average rate in the light curve, assuming a Poisson probability density function $P(x)$. A potential burst is flagged if $P(x)<0.01/N$, where $N$ is the number of bins per light curve (typically $100/t_{\rm r}$). The flagged bins are removed and the procedure is repeated until no more bins are found. We chose $t_{\rm r}$ to be 8~ms, 32~ms, and 128~ms to sample the full range of magnetar burst durations. We confirm our GBM search results with a Bayesian blocks search technique as detailed in \citet{lin20ApJ}. We only inspect potential bursts that were found in two or more detectors; single detector enhancements are typically due to particle background rather than an astrophysical event. Moreover, we rejected all bursts when \src\ was behind the earth for \fermi, and those that were registered only in NaI detectors with an angle $>60^\circ$ from the source since their effective area drops significantly at such large observing angles. The source identification process, however, is complicated by another magnetar, SGR~J1555$-$5402, which is 40\arcmin\ from \src\ and was also active at this time. This separation is not distinguishable by GBM, but is easily resolved in BAT images.

We find six candidate bursts that may originate from \src\ between 2022 March 16 and April 7, including one that corresponds to a GBM trigger on 2022 April 3 (trigger 220403928). 
BAT detected a count rate increase in two cases, in addition to the short-burst detected on 2022~April~7.
One of these was identified as the aforementioned SGR~J1555$-$5402, and the other coincidence has a negative value (consistent with zero) at the \src\ location and excludes it as the source of the rate increase. 
Hence, the apparent lack of any bursting activity associated with \src\ over the course of three weeks following the last radio detection suggests that the high energy outburst genuinely commenced later. 
However, we cannot establish the precise date of the outburst, and tentatively place it sometime in early April. 
Future continuous monitoring of X-ray bright radio-loud magnetars is crucial for revealing the dynamic between the radio and X-ray emission at the start of an outburst.

\section{Discussion} \label{sec:disc}

Prior to the 2009 outburst, \src\ was considered to be a so-called transient magnetar, similar to XTE~J1810$-$197 \citep{Ibrahim2004}, with a historical quiescent (absorbed, 0.1--10\,keV) flux of (0.3--2.1)$\times 10^{-12}$\,erg\,s$^{-1}$\,cm$^{-2}$ \citep{Gelfand2007}. However, the 2009 outburst was associated with a transition of \src\ into a persistent magnetar state where it exhibits a significantly higher steady-state X-ray flux than the pre-2006 quiescent state \citep{CotiZelati2020}. On account of this increased persistent emission, the 2022 outburst appearing to be both less intense and shorter lived than the 2008 outburst is likely an artifact of the much higher baseline X-ray flux. Indeed the maximum 0.5--10\,keV flux of $(6.0 \pm 0.4) \times 10^{-11}$\,erg\,s$^{-1}$\,cm$^{-2}$ is comparable to the observed peak flux of $\sim 6 \times 10^{-11}$\,erg\,s$^{-1}$\,cm$^{-2}$ during the 2008 outburst \citep{Israel2010}. Had this previous outburst occurred more recently, the X-ray flux of \src\ would have fallen below the current baseline of $\sim 1.7 \times 10^{-11}$\,erg\,s$^{-1}$\,cm$^{-2}$ after a similarly short 10--20\,day timescale. Were it not for our triggered \nicer\ observations, the enhanced X-ray emission from this outburst may have been missed entirely. A smaller outburst may not have even produced a detectable X-ray enhancement.

Magnetar X-ray outbursts are generally attributed to crustal shifts due to stresses on the surface from internal B-field decay. These shifts could also either modify or enhance toroidal twists in the morphology of external magnetic field lines \citep[see][and references therein]{turolla15RPPh}. The soft X-ray thermal emission arises due to either internal energy deposition in the outer crust \citep{beloborodov16ApJ, akgun18MNRAS, lander2019MNRAS}, bombardment of the surface by accelerated particles in the twisted magnetospheric field loops \citep{Beloborodov2009}, mechanical stresses slowly building in the crust (see \citealt{Lander2015, Kojima2022} and references therein) or a combination of all three. In this model, a more complex pulse shape is expected during the outburst due to the emergence of new hot spots on the surface \citep[e.g.,][]{Archibald2017a,younes22ApJL18301}, commensurate with the 3--6 keV pulse profile morphology displayed in Fig.~\ref{Xprof}. As the outburst abates, the source surface heat map resorts back to its long-term, quasi-sinusoidal configuration.  If surface bombardment is active, the evolving surface temperature profile likely is accompanied by a migration of field line footprints associated with the twisting and untwisting of the magnetospheric field.

The hard X-ray tail above 10 keV is believed to arise from resonant inverse Compton scattering (RICS) of the soft surface thermal photons by the relativistic electrons accelerated in twisted loops \citep{baring07ApSS,nobili08MNRAS,wadiasingh2018ApJ}. Figs.~\ref{specFig} and~\ref{Xprof} clearly display a lack of any variability in the hard X-ray emission concurrent with the elevated soft thermal emission during this outburst of \src.  This disconnect between soft and hard X-ray bands is to be expected.  Detailed models of RICS emission generally favor a production of the signal at quasi-equatorial locales of modest altitudes, where scattering kinematics between the electrons and the soft X-ray photons from the surface readily accesses the cyclotron resonance in the cross section \citep{wadiasingh2018ApJ}.  Thus changes in field line geometry near their footprints (e.g. due to a localized crustal stress) and moreover near the polar regions likely have only a small impact on the RICS emission, as do alterations of the surface hotspot size and temperature.  Evidence for some separation of magnetic co-latitudes for hard X-ray signals and the surface soft X-ray ones may be suggested by the modest 0.1-0.3 phase lag of the peak between the 3--6 keV band and the 10--25 keV one in Fig.~\ref{Xprof}.  Thus changes in pulsed radio emission emanating from in or near the open field region, might naturally not be accompanied by modifications to the hard X-ray signal.

Generally, pulsed radio emission from magnetars is closely linked to the outburst mechanism: it switches on following the onset of an outburst before gradually fading into radio silence as the neutron star settles into X-ray quiescence (e.g., \citealt{Camilo2016, Scholz2017}, but see \citealt{zhu20ATel14084,younes22arXiv221011518Y}). Unlike other radio-loud magnetars, persistent radio pulses have been detected from \src\ since the quasi-stabilization of its X-ray flux to the current high state in $\sim$2013 (\citealt{CotiZelati2020}; Camilo et al., in prep.).
As shown in Figure~\ref{fig:flux}, radio pulses were detected up until sometime between 2022 March 16 and 26, and were not seen again until April 19. 
An initial disappearance and reactivation of the radio emission of \src\ was also reported following the 2009 outburst \citep{Camilo2009ATel, Burgay2009ATel}.

The 2016 magnetar-like outburst of the high magnetic field strength pulsar PSR~J1119$-$6127 was also associated with a shutdown in the persistent radio emission, which was re-detected around two weeks after the initial X-ray outburst \citep{Dai2018}. 
Moreover, simultaneous X-ray and radio observations of PSR~J1119$-$6127 revealed that magnetar-like short bursts were associated with a sudden disappearance of pulsed radio emission \citep{Archibald2017b}.
This correlation was interpreted as a burst-induced quenching of the radio emission arising from electron-positron pairs leaking from the fireball trapped above the surface into the particle acceleration gaps, thereby increasing the plasma density within these regions and suppressing the acceleration of charges that precipitate radio emission.
A similar phenomenon would provide a natural explanation for the disappearance of radio pulses from \src\ if it occurred after the onset of the high-energy outburst. 
Yet, the dearth of \fermi-GBM and \swift-BAT burst candidates up to early April 2022 implies a delayed onset of the high energy outburst with respect to the disappearance of radio pulses.
An earlier onset necessitates an unlikely scenario in which all short-bursts emitted over an extended period of time were missed by all large FoV hard X-ray and gamma-ray monitors.
Alternatively, a slow undetected rise in high-energy activity could explain the observed delay between the radio disappearance and delayed onset of short burst detections. 
Such behavior was previously seen in the 2002 outburst of 1E~1048.1$-$5937 \citep{Archibald2020}.

Small-scale twisting of the magnetosphere from crustal fracturing and motion of the magnetic footprints can lead to a decrease or complete failure of the charged particle currents responsible for producing the pulsed radio emission in addition to variations in the optical depth of the magnetosphere \citep{Beloborodov2009, Timokhin2010, Hu2022}. 
Regardless of the exact underlying physics, any alteration of the magnetosphere must have been short lived, as we did not detect any substantial change in the average radio pulse profile shape or polarization properties of \src\ after re-activation.
This is markedly different to the behavior of other radio-loud magnetars, which typically display a broad range of profile variations and a marked increase in radio flux in the immediate aftermath of an outburst (e.g., \citealt{Scholz2017, Dai2018, Dai2019}). 
Future observations of similar high-energy outbursts both in \src\ and other highly magnetized pulsars will shed further light on the radio-quenching mechanism.

Our joint radio and X-ray observations revealed that the 2022 outburst resulted in a significant spin-up event in the timing of \src, which bears a strong resemblance to a pulsar glitch.
The gap in our timing coverage due to the disappearance of pulsed radio emission, low observing cadence of \swift\ and beginning of our triggered \nicer\ observations means the exact epoch of the event is unknown, but is constrained to between MJD 59661 to 59672 (March 23 -- April 3, 2022).
Glitch-like timing anomalies and subsequent rapid changes in $\dot{\nu}$ are common feature of the post-outburst timing of magnetars (e.g., \citealt{dib14ApJ, Archibald2020}), the latter of which are often ascribed to torque variations imparted by twists in the magnetosphere \citep{Thompson2002}.
Unlike rotation-powered pulsars, the majority of glitches in magnetars are associated with radiative outbursts that are believed to be caused by either starquakes \citep{Thompson1995} or magnetic re-connection events \citep{Lyutikov2002}.
In the starquake model, the gradual untwisting of the internal magnetic field induces an enormous strain on the rigid outer crust of the neutron star.
Eventually the tensile strength of the crust is overcome, leading to a sudden plastic deformation or even fracturing of a small region of the crust.
This triggers an outward exchange of angular momentum from the superfluid core to the crust via unpinning of vortices at the crust-core boundary \citep{Ruderman1998}.
Modelling the timing anomaly in \src\ as a glitch, we obtained permanent step-changes in $\nu$ and $\dot{\nu}$ of $0.2(1)$\,$\mu$Hz and $-2.4(1) \times 10^{-12}$\,s$^{-2}$ respectively.
The permanent change in spin-frequency is small when compared to most glitches in magnetars (see Figure 2 of \citealt{Fuentes2017}).
On the other hand, the increase in spin-down rate is among the largest recorded in any pulsar \citep{Lower2021b, Basu2022}.

Following the glitch, we also detected a secular increase in the spin-down rate that continued up until at least our last \nicer\ observation.
We inferred a $\ddot{\nu}$ of $-1.95(5) \times 10^{-19}$\,s$^{-3}$ from a simple linear fit to the $\dot{\nu}$ timeseries and $-2.0(1) \times 10^{-19}$\,s$^{-3}$ from our glitch fit in Section~\ref{timSec}.
This corresponds to a net increase in $\dot{\nu}$ of $-2.54 \times10^{-12}$\,s$^{-2}$ over the 147-days of post-outburst timing.
A similar linear trend was observed following the 2008 outburst \citep{Israel2010}, where the resulting $\ddot{\nu} = -4.8(9) \times 10^{-18}$\,s$^{-3}$ was an order of magnitude larger than we observed after the 2022 outburst. 
Secular increases in spin-down rate are a common feature of outbursts in magnetars and high B-field pulsars and are often attributed to the magnetosphere becoming increasingly twisted up to a maximum value after a few 10's to 100's of days \citep{Thompson2002, Beloborodov2009}.
After reaching the maximum twist, the field bundle then begins to unwind, causing a turnover in the spin-down evolution as the magnetar relaxes back toward the pre-outburst state.
A global alteration of the twist can be ruled out in this 2022 event given the lack of hard X-ray evolution, indicating that either a small-scale local twist in bundles of magnetic field lines near the active pole or transient particle winds may be responsible for the secular increase in $\dot{\nu}$ \citep{Harding1999}.

Spin-down rates peak anywhere between $-10^{-13}$ -- $10^{-11}$\,s$^{-2}$ depending on the strength of the outburst.
Given the lack of a turnover in the post-outburst $\dot{\nu}$ evolution of \src, it is likely the maximum twist imparted by the 2022 outburst or peak in the outflowing particle flux has yet to be reached.
The recovery back to the initial spin-down state varies between magnetars and from one outburst to the next.
Occasionally, the initial turnover is followed by an extended period of `noisy' spin-down evolution, including quasi-periodic variations such as those seen in 1E~1048.0$-$5937 \citep{Archibald2015} and Swift~J1818.0$-$1607 \citep{Rajwade2022}. 
These variations are reminiscent of the spin-down state-switching process seen among rotation-powered pulsars \citep{Lyne2010}, and often not associated with any changes in high-energy emission.
Such behavior remains to be seen in the current post-outburst timing of \src.

\section{Conclusions}

We have presented the detection of a new outburst of \src\ that began in 2022 April. The outburst was marked by an increased soft X-ray emission and the disappearance of pulsed radio emission, both of which recovered back to their pre-outburst properties $\sim$3-weeks later. 
The disappearance of the radio pulses prior to the outburst onset is surprising, and could point to an undetected slow rise in high energy activity. 
Such behavior could arise from small-scale crustal fracturing, where variations in the magnetospheric currents and optical depth resulted in a quenching of the radio emission mechanism.
Interestingly, we find no change to the hard non-thermal tail concurrent to the soft X-ray and radio variability. Our joint \parkes/\nicer\ timing measurements revealed the presence of a glitch-like timing event in the rotation of the magnetar.
A significant instantaneous increase in spin-down rate was also measured, which continued to increase at a rate of $-2.75 \times 10^{-19}$\,s$^{-3}$ over the first 147-days after the outburst.
Small-scale shifts in magnetospheric geometry in the lead up to the main quake may have been responsible for the quenching of the pulsed radio emission.
These results highlight the importance of conducting high-cadence monitoring of radio-loud magnetars such as \src\ at both radio and X-ray frequencies. Future outbursts with greater coverage around the initial trigger time may be able to resolve the radio quenching, delayed rise in spin-frequency, and initial increase X-ray flux.
Such observations would provide an even clearer picture of the magnetar outburst mechanism.
Given some similarities between this latest event and the 2008 outburst, it is possible that \src\ has entered a new state of increased activity.
In this case, the 2022 outburst may therefore be a precursor to additional outbursts over the coming years.

\begin{acknowledgments}
We thank George Hobbs and Rahul Sengar for generously giving us some of their Parkes observing time, and a special thanks to James Green and Phil Edwards for granting Parkes director discretionary time. 
We also thank the anonymous referee for their helpful comments.
The Parkes radio telescope (\textsl{Murriyang}) is part of the Australia Telescope National Facility (\href{https://ror.org/05qajvd42}{https://ror.org/05qajvd42}) which is funded by the Australian Government for operation as a National Facility managed by CSIRO. 
We acknowledge the Wiradjuri people as the traditional owners of the Observatory site. 
Radio data reduction and analysis were performed on the OzSTAR national HPC facility, which is funded by Swinburne University of Technology and the National Collaborative Research Infrastructure Strategy (NCRIS). 
A portion of this work was supported by NASA through the NICER mission and the Astrophysics Explorers Program. 
G.~Y. research is supported by an appointment to the NASA Postdoctoral Program at the Goddard Space Flight Center, administered by Oak Ridge Associated Universities under contract with NASA. 
L.~D. is supported by the Australian Research Council Centre of Excellence CE170100004 (OzGrav). M.~G.~B. acknowledges the generous support of the National Science Foundation through grant AST-1813649.  Z.~W. acknowledges support by NASA under award number 80GSFC21M0002. S.~G. acknowledges the support of the National Centre for Space Studies (CNES).
\end{acknowledgments}

\software{{\tt dspsr} \citep{vanStraten2011}, {\tt HEAsoft} (v6.30.1; \citealt{HEASARC2014}), {\tt psrchive} \citep{Hotan2004, vanStraten2012}, {\tt NuSTARDAS} (v2.1.2), {\tt DM\_phase} \citep{Seymour2019}, {\tt PRESTO} \citep{Ransom2011}, {\tt RMNest} \citep{Lower2022}, {\tt tempo2} \citep{Hobbs2006, Edwards2006}, {\tt TempoNest} \citep{Lentati2014}, {\tt XSPEC} (v12.12.1; \citealt{Arnaud1996})}

\bibliography{main}{}
\bibliographystyle{aasjournal}

\end{document}